\documentclass[12pt]{article}
\usepackage{authblk,graphicx,caption,figsize}
\makeatletter
\newcommand{\singlespacing}{\let\CS=\@currsize\renewcommand{\baselinestretch}{1.0}\tiny\CS}
\newcommand{\doublespacing}{\let\CS=\@currsize\renewcommand{\baselinestretch}{1.25}\tiny\CS}

\begin{document}
\title{Analyzing Non-Extensivity of $\eta$-spectra in Relativistic Heavy Ion Collisions at $\sqrt{s_{NN}}=200$ GeV}

\author{Bhaskar De$^1$\thanks{e-mail: de$\_$bhaskar@yahoo.com}, Goutam
Sau$^2$\thanks{e-mail: gautamsau@yahoo.co.in}, S. K.
Biswas$^3$\thanks{e-mail: sunil$\_$biswas2004@yahoo.com}, S.
Bhattacharyya$^4$\thanks{e-mail: bsubrata@www.isical.ac.in} \&  P. Guptaroy$^5$\thanks{e-mail: gpradeepta@rediffmail.com} }

\affil{$^1$Department of Physics, Moulana Azad College\\
8, Rafi Ahmed Kidwai Road, Kolkata - 700013, India.\\
$^2$ Beramara Ramchandrapur High School,\\
South 24-Pgs,743609(WB),India.\\
$^3$ West Kodalia Adarsha Siksha Sadan, New Barrackpore,\\
Kolkata-700131, India.\\
$^4$Physics and Applied Mathematics Unit$(PAMU)$,\\
Indian Statistical Institute, Kolkata - 700108, India.\\
$^5$Department of Physics, Raghunathpur College\\
Raghunathpur-723133, Purulia, India.}

\date{}
\maketitle

\begin{abstract}
The transverse momentum spectra of secondary $\eta$ particles produced in $P+P$, $D+Au$ and $Au+Au$ interactions at $\sqrt{s_{NN}}=200$ GeV at different centralities have been studied in the light of a non-extensive thermodynamical approach. The results and the possible thermodynamical insights, thus obtained, about the hadronizing process have also
been discussed in detail.
\ \\

\noindent {\bf Keywords} \ : \ Relativistic Heavy Ion Collision, Inclusive Cross Section.\\

\noindent {\bf PACS Nos.} \ : \ 25.75.-q, 13.60.Hb.
\end{abstract}
\newpage
\doublespacing
\section{Introduction}
The multiparticle production in relativistic heavy ion collision
provides important information about the hadronizing process in the
aftermath of any high energy collision phenomenon. Various
phenomenological/mathematical models have, so far, been utilised to
analyse the available experimental data to understand the different
characteristics and dynamical aspects of this process. Amongst the
various existing models, applicability of Tsallis non-extensive
thermodynamical approach\cite{Tsallis1,Tsallis2,Tsallis3,Prato1,Tsallis4,Tsallis5,Tsallis6}, which has already left it's
marked impression in other branches of physics, has recently gained
pace in understanding mainly the statistical as well as the
thermodynamical characteristics of the production of various
secondaries in high energy nuclear interactions at
 different energies\cite{Beck1,Beck2,Beck3,Wilk1,Wilk2,Wilk3,Wilk4,Wilk5,Osada1,Wilk6,Biro1,Biro2,Biro3,Biyajima1,Biyajima2,Wilk7,Wilk8,Alberico1,Lavagno1,Kaniadakis,Kodama}.
We have taken up, here in this paper, the task of analysing the
transverse momentum spectra of one of the heavier secondaries,
$\eta$-mesons, produced in different nuclear reactions at RHIC-BNL
in the light of Tsallis non-extensive thermodynamics.  This is  in
continuation of our previous work\cite{De1}, wherein transverse
momentum spectra for the secondaries like $\pi^+$, $\pi^-$, $P$ and
$\bar{P}$ were treated with the same approach. As in our previous
work\cite{De1}, here also we have confined ourselves only to a
particular interaction
energy($\sqrt{s_{NN}}=200$ GeV), so that a
comparative study can be made among various nucleus-nucleus systems.
Our mention of the $P+P$-interactions in the same breath and bracket
with the host of nuclear reactions is to be viewed somewhat loosely;
this was done with our eyes cast on the same interaction energy
alone. This is specially so when the `centrality'-aspects are
touched upon and dealt with. In our actual work-procedure we have
been attentive to the particularities of each interaction on a
case-to-case basis. In the recent past, Biro et al\cite{Biro2,Biro3}
analyzed successfully transverse momentum spectra of different
identified hadrons, including $\eta$-mesons, produced at RHIC
energies, on the basis of both the non-extensive statistical
approach and the very essence of quark-coalescence model. However,
the study was, probably by choice, confined to the minimum bias
$Au+Au$ collisions in central rapidity region. Our present study,
however, takes into account (i) the effect of entire longitudinal
momentum($p_L$) range, (ii) the role of the same in different
nuclear collisions($Au+Au$, $D+Au$ and $P+P$) and (iii) the impact
of centrality dependent windows.
\par
The $\eta$ mesons are the heavier species among the various mesonic secondaries produced in high energy nuclear reactions.  The formation of such heavy partonic bound states and their interactions in nuclear matter under different conditions throws light on the nature of the possible hadronic phase transition to a partonic state, with a greater degrees of freedom, formed immediate after the nuclear collisions at relativistic energies.  The transverse momentum spectra of such secondaries at different centralities provides
significant clues on the mechanism of the formation of such bound states, on their partonic constituents as well as on the thermal and dynamical characteristics of the
regions in the partonic matter where they emerged from.
\par
The work is organized as follows: Section 2 presents an outline of
the hadronization process. In Section 3 we give some very brief
sketch of the nonextensive statistics and the main working formula
to be used in our work here.  The results are reported in next
section(Section 4) with some specific observations made. And the
last section is preserved for our conclusions.
\section{Non-extensivity and Hadronization Process: An Outline}
The evolution of the partonic system created in RHIC-BNL experiments
is generally believed by a large section of the theoretical
physicists to be best described by hydrodynamics of an ``almost
ideal fluid". This approach really gives a fair description of data
on the particle transverse momentum spectra and asymmetry of the
transverse flow. However, the prospect of this theory and
methodology is, at the same time, riddled with and marred by two
very serious problems: (i) the `HBT puzzle' and (ii) the `puzzle of
early thermalization'\cite{Bialas,Ryblewski} which we are not going
to elaborate here in detail. Besides this, the proposed microscopic
mechanisms lack in the detailed knowledge of fragmentation and
recombination etc. These deficiencies have prompted the physicists
to turn to many different models, often from various different
angles, of which the non-extensive thermodynamic approach is an
effective one which is, thus, essentially a thermal model. This
model assumes the formation of a system which is thermally and
chemically in a near-equilibrium state in the hadronic phase and is
characterised, in the main, by two thermodynamic
variables(observables), for the hadronic phase. These two are
`temperature' and `non-extensive entropy' of the system attained
finally. The deconfined period of the time evolution dominated by
the constituent partons(quarks and gluons) remains hidden: full
equilibration generally washes out and destroys large amount of
information about the early deconfined phase. Successes of thermal
statistical models are shaped and conditioned by the blackening out
of information on some intermediate states. And this is an accepted
reality.
\par
In the case of full thermal and chemical equilibrium relativistic
statistical distributions are generally based on exponential nature
of the spectra for the $p_T$- distribution of the produced hadronic
secondaries. But, observationally experimental data at SPS and RHIC
energies reveal strongly the traits of non-exponential and power-law
nature at high-$p_T$ which is also supported by the dictates of the
perturbative QCD\cite{Fermi,Heisenberg,Hagedorn1}. So, for the
overlap or the transitional regions of the
$p_T$-values(low-to-high), a stationary distribution of strongly
interacting hadron gas or quark-matter in a finite volume could be
considered to have such a distribution with dual nature. In fact,
Tsallis distributions\footnote{However, the coinage `Tsallis distribution' is not fully justified, because Vilfredo Pareto\cite{Pareto} used much earlier and long ago a power-law probability distribution found in a large number of real-world situations, especially in the field of Economics. And beyond the borders of economics, it is generally referred to as the Bradford distribution, though the high energy physicists, since 1970s, prefer to term this as a ``cut-power-law distributions". In the context of non-extensive thermodynamics alone, this is normally familiar as Tsallis distribution.}, which we have resorted to, satisfy such a
criterion\cite{Cleymans} .

\section{Nonextensive Statistics and Transverse Momentum Spectra}

The nonextensive entropy by Tsallis generalized statistics is
given by\cite{Tsallis1},

\begin{equation}
S_q ~ = ~ \frac{1}{q-1} ~ ( ~ 1 ~ - ~ \sum_i ~ p_i^q ~)
\end{equation}

where $p_i$ are probabilities associated with the microstates of a
physical system with normalization $ \sum_i ~ p_i=1$ and $q$ is the nonextensivity parameter. For $q ~
\rightarrow ~ 1$, eqn.(1) gives the ordinary Boltzmann-Gibbs
entropy

\begin{equation}
S ~ = ~ - ~ \sum_i ~ p_i ~ \ln{p_i}
\end{equation}

The generalized statistics of Tsallis is not only applicable to an
equilibrium system, but also to nonequilibrium systems with
stationary states\cite{Beck2}. As the name `nonextensive' implies,
these entropies are not additive for independent systems. For a
system of $N$ independent particles, where particle 1 is in energy state
$\epsilon_{i_1}$, particle 2 in energy state $\epsilon_{i_2}$, and so on,
the Hamiltonian of the system in nonextensive approach is given by\cite{Beck2},

\begin{equation}
H(i_1,i_2,...,i_N) ~ = ~ \sum_j ~ \epsilon_{i_j} ~ + ~ (q-1)\beta ~ \sum_{j,k} ~
\epsilon_{i_j} \epsilon_{i_k} ~ + ~ (q-1)^2\beta^2 ~ \sum_{j,k,l}
~ \epsilon_{i_j} \epsilon_{i_k} \epsilon_{i_l} ~ + ~ \ldots
\end{equation}

where  $\beta=1/T$ is the inverse temperature variable.
The above equation clearly indicates that for a non extensive
system the total energy is not the sum of the single-particle
energies.
\par
The energy associated with a particle, which is a hadron in the present context,  denoted by $j$ in a momentum
state $i$ in a fireball produced in a high energy nuclear
collision is given by,

\begin{equation}
\epsilon_{ij} ~ = ~ \sqrt{\textbf{p}_i^2 ~ + ~ m_j^2}
\end{equation}

The nonextensive Boltzmann factor is defined as\cite{Beck2}

\begin{equation}
x_{ij} ~ = ~ (1 ~ + ~ (q-1)\beta \epsilon_{ij})^{-q/(q-1)}
\end{equation}

with $q\rightarrow1$, the above equation approaches the ordinary
Boltzmann factor $e^{-\beta\epsilon_{ij}}$. If $\nu_{ij}$ denotes
the number of particles of type $j$ in momentum state $i$, the
generalized grand canonical partition function is given by,

\begin{equation}
Z ~ = ~ \sum_{(\nu)} \prod_{ij} x_{ij}^{\nu_{ij}}
\end{equation}

The average occupation number of a particle of species $j$ in the
momentum state $i$ can be written as\cite{Beck2}

\begin{equation}
{\bar{\nu}_{ij}}= x_{ij}\frac{\partial}{\partial x_{ij}} \log{Z} ~
= ~ \frac{1}{(1+(q-1)\beta\epsilon_{ij})^{q/(q-1)}\pm 1}
\end{equation}

where $-$ sign is for bosons and the $+$ sign is for fermions.\\
The probability of observation of a particle of mass $m_0$ in a certain
momentum state can be obtained by multiplying the average occupation number with
the available volume in momentum space\cite{Beck2}. The infinitesimal volume in momentum space
is given by

\begin{equation}
dp_x ~ dp_y ~ dp_z ~ = ~ dp_L ~ p_T ~ sin\theta ~ dp_T ~ d\theta
\end{equation}

where $p_T=\sqrt{p_x^2+p_y^2}$ is the transverse momentum and $p_z=p_L$ is the longitudinal one.
Hence, the probability density $w(p_T)$ of transverse momenta is obtained by integrating over
all $\theta$ and $p_L$:

\begin{equation}
w(p_T) ~ = ~ const. ~ p_T ~ \int_{-\infty}^{+\infty} ~ dp_L
\frac{1}{(1+(q-1)\beta\sqrt{p_T^2+p_L^2+m_0^2})^{q/(q-1)} \pm 1}
\end{equation}
Since, the integrand in the right hand side of the above equation
is an even function of $p_L$, one can easily write,
\begin{equation}
w(p_T) ~ = ~ const. ~ p_T ~ \int_0^{+\infty} ~ dp_L
\frac{1}{(1+(q-1)\beta\sqrt{p_T^2+p_L^2+m_0^2})^{q/(q-1)} \pm 1}
\end{equation}

Since the temperature $T$, which will be called here as Hagedorn
temperature and will be denoted as $T_{eff}$, is quite small, i.e.,
$\beta\sqrt{p_T^2+p_L^2+m_0^2}\gg 1$, one can neglect $\pm 1$ in
the denominator of the previous equation. Hence, the differential
cross section, which is proportional to $w(p_T)$, can be expressed
as,

\begin{equation}
\frac{1}{\sigma}\frac{1}{p_T} \frac{d\sigma}{dp_T}~ \propto ~
\int_0^\infty dp_L
(1+\frac{(q-1)}{T_{eff}}\sqrt{m_0^2+p_T^2+p_L^2})^{-q/(q-1)}
\end{equation}

or, in terms of the transverse mass($m_T^2= m_0^2+p_T^2$) of the detected secondary,

\begin{equation}
\frac{1}{\sigma}\frac{1}{m_T} \frac{d\sigma}{dm_T}~ \propto ~
\int_0^\infty dp_L
(1+\frac{(q-1)}{T_{eff}}\sqrt{m_T^2+p_L^2})^{-q/(q-1)}
\end{equation}

If we put $x=\frac{p_L}{T_{eff}}$ and $u=\frac{m_T}{T_{eff}}$, the above equation can be written as,

\begin{equation}
\frac{1}{\sigma} \frac{d\sigma}{dm_T}~ \propto ~
u\int_0^\infty dx
( ~ 1+(q-1)u\sqrt{1+\frac{x^2}{u^2}} ~ )^{-q/(q-1)}
\end{equation}

For large $x$ the integrand is small and it's contribution to the
integration can be ignored\cite{Beck1}. Hence, for large $u$ and
small $x$ we can assume, $\sqrt{1+\frac{x^2}{u^2}} \approx 1+
\frac{x^2}{2u^2}$. The inclusion of this assumption in the above
equation yields

\begin{equation}
\frac{1}{\sigma} \frac{d\sigma}{dm_T}~ \propto ~
u(1+(q-1)u)^{-q/(q-1)} \int_0^\infty dx
( ~ 1+\frac{(q-1)x^2}{2u(1+(q-1)u)} ~ )^{-q/(q-1)}
\end{equation}
This form of equation is in accordance with the right hand side of
eqn(28) given in \cite{Beck1}. So, one can write, following the
eqn(32) of the same reference,
\begin{equation}
\frac{1}{\sigma} \frac{d\sigma}{dm_T}~ \propto ~
u^{3/2} (1+(q-1)u)^{-\frac{q}{q-1}+\frac{1}{2}}
\end{equation}
or,
\begin{equation}
\frac{1}{\sigma} \frac{d\sigma}{dm_T} ~ = ~ c_1 ~
u^{3/2} (1+(q-1)u)^{-\frac{q}{q-1}+\frac{1}{2}}
\end{equation}

where $c_1$ is a normalization constant.
\par
The average multiplicity of the detected secondary in the given rapidity region can be obtained by the relationship
\begin{equation}
<N> = \frac{1}{\sigma} \int_{m_0}^\infty \frac{d\sigma}{dm_T} ~ dm_T ~ = ~ c_1 ~
\int_{m_0}^\infty u^{3/2} (1+(q-1)u)^{-\frac{q}{q-1}+\frac{1}{2}} dm_T
\end{equation}

Hence, the constant $c_1$ can be expressed in terms of $<N>$ by the relationship

\begin{equation}
c_1 = \frac{<N>}{T_{eff}
\int_{u_0}^\infty u^{3/2} (1+(q-1)u)^{-\frac{q}{q-1}+\frac{1}{2}} du}
\end{equation}
where $u_0=\frac{m_0}{T_{eff}}$.
\par
Combining eqn(14) and eqn(16) one can write
\begin{equation}
\frac{1}{\sigma} \frac{d\sigma}{dm_T} ~ = ~ \frac{<N>}{T_{eff}
\int_{u_0}^\infty u^{3/2} (1+(q-1)u)^{-\frac{q}{q-1}+\frac{1}{2}} du} ~
u^{3/2} (1+(q-1)u)^{-\frac{q}{q-1}+\frac{1}{2}}
\end{equation}
Since,
\begin{equation}
\frac{1}{\sigma} \frac{1}{m_T} \frac{d\sigma}{dm_T} ~ = ~ \frac{1}{\sigma} \frac{1}{p_T} \frac{d\sigma}{dp_T}
~ = ~ \frac{dN}{p_T ~ dp_t},
\end{equation}
 we can write
\begin{equation}
\frac{dN}{p_T ~ dp_t} ~ = ~ \frac{<N>}{T^2_{eff}
\int_{u_0}^\infty u^{3/2} (1+(q-1)u)^{-\frac{q}{q-1}+\frac{1}{2}} du} ~
u^{1/2} (1+(q-1)u)^{-\frac{q}{q-1}+\frac{1}{2}}
\end{equation}
Equation(21) provides the working formula for the present analysis.

\section{Results}
The working formula was applied, in it's present form, to obtain a fit to the
data on $\eta$ production in $P+P$ collision[Fig.1(a)] at $\sqrt{s_{NN}}=200$ GeV and
the corresponding parameter-values are given in Table-1, where $n_o$ denotes the average
multiplicity of $\eta$ produced in $P+P$ interaction.
\par
However, when the same formula was employed to analyse data from nucleus-nucleus collisions
like $D+Au$ and $Au+Au$ collisions, the parameters, which had been set free for $P+P$ collisions,
were constrained by the following relationships\cite{Wilk8}:

\begin{equation}
T_{eff}=T_0(1-c(q-1))
\end{equation}

\begin{equation}
\frac{<N>-n_0N_{part}}{<N>}=c(q-1)
\end{equation}

with $c= - \frac{\phi}{D c_p\rho T_0}$ where $D$, $c_p$, $\rho$,
$T_0$ are respectively the strength of the temperature fluctuations,
the specific heat under constant pressure, density, the temperature
of the hadronizing system when it is in thermal equilibrium($q=1$)
and $N_{part}$ is the number of participant nucleons. Eqn.(22)
describes the fluctuation in temperature where it is assumed that
the effective temperature $T_{eff}$ is the outcome of two
simultaneous processes: (i) the fluctuation of the temperature
around $T_0$ due to a stochastic process in any selected region of
the system and (ii) some energy transfer between the selected region
and the rest of the system, denoted by $\phi$\cite{Wilk8}. It is
absolutely uncertain whether such energy-transfers could/should be
invariably linked up with flow-velocity(normally denoted in the
hydrodynamical model-texts as `$u$'). So, for the sake of
calculational simplicity and correctness we assume the factor
$\phi$, for the present, to be independent of any flow-velocity. The
fluctuation in multiplicity is described by eqn.(23). The assumption
behind this relationship is that if N-particles are distributed in
energy according to Tsallis non-extensive distribution, then their
multiplicity will obey Negative-Binomial distribution\cite{Wilk8}.
\par
The fits for $D+Au$ and $Au+Au$ collisions obtained on the basis of
equation(21) alongwith the constraints given in eqn(22)-(23) are
depicted in Fig.1(b)-Fig.1(d). The values of various parameters
obtained from the fits are given in Table-2. The values of $T_{eff}$
and $q$ calculated from the fitted parameters are given in tabular
form in Table-3 and in graphical format in Fig.2 as a function of
participant nucleons. Fig.3 depicts graphically the behaviour of
average  multiplicity of $\eta$-mesons produced per pair of
participant nucleons as a function of $N_{part}$. The normalized
values of $<N>$ for different nucleus-nucleus collisions at
$\sqrt{s_{NN}}=200$ GeV remain almost constant with respect to
$N_{part}$ which indicates the linear dependence of $<N>$ on
$N_{part}$, and hence on the system size.
\par
The obtained values of $q$ and $T_{eff}$, on the average, show
centrality-dependences for almost all the collisions. The values of
$T_0$ and $c$ obtained from different fits show almost constant
behaviour according to expectation and the average values of them
are found to be 155 MeV and 1.88 respectively. The performance of
the present non-extensive approach vis-a-vis the studied
experimental data could be rated to be moderately satisfactory, as
is seen from the $\chi^2/ndf$-values given in the last column of
Table-1.

\section{Discussion and Conclusions}
The centrality dependences of the parameters $q$ and $T_{eff}$ are
quite clear from the Fig.2. The systematic trend of the
non-extensive parameter $q$, in general, is that it decreases with
increase in the number of participant nucleons, i.e. with the
increase in centrality. On the other hand, the Hagedorn temperature
$T_{eff}$ exhibits completely a different trend, i.e., it increases
with the increase in centrality. The obtained values of $q$ and
$T_{eff}$ for $D+Au$, and $Au+Au$ collisions are in agreement with
these tendencies.
\par
The contrasting behaviour of $q$ and $T_{eff}$ with respect to
$N_{part}$ is in accord with the spirit and content of the
non-extensive statistics. $q$ is related with the fluctuations in
temperature\cite{Wilk1,Biyajima1}. The high value of $q$ means a
high fluctuation in temperature and the system is not in the
neighbourhood of its thermal equilibrium. The number of binary
collisions in a system with lesser number of participant nucleons is
quite low. Hence, the probability of mutual exchange of transverse
momenta among the interacting partons in a system involving small
$N_{part}$ is also quite low, which, in turn, keeps the system far
away from its thermal equilibrium. On the other hand, an appreciable
increment in the number of binary collisions is observed in a system
possessing a large number of  participant nucleons; and consequently
the system can reach quickly to its thermal equilibrium or in the
close vicinity of it. So it is quite natural that the fits, obtained
with the non-extensive approach, will exhibit a gradual decrement in
the values of $q$ and increment in $T_{eff}$, as one goes from
central to peripheral region or from a system with larger $N_{part}$
to that with smaller ones.
\par
$q$ and $T_{eff}$ change rapidly when $N_{part}$ lies in the range
$2\leq N_{part}\leq 20$. The change is rather slow for high
$N_{part}$. The dependence of $q$ and $T_{eff}$ on $N_{part}$ was
observed quite strong over the entire $N_{part}$-region while
studying charged pions and proton-antiprotons\cite{De1} produced in
nuclear collisions at $\sqrt{s_{NN}}=200$ GeV. This fact is also
reflected from the values of $T_0$ and $c$. The values of $T_0$ for
charged pions and proton-antiprotons, obtained by Wilk et
al\cite{Wilk7} on the basis of our previous venture\cite{De1}, are
220 MeV and 360 MeV respectively while that of $c$ are 5.7 and 9.4
respectively. But, the values of $T_0$ and $c$, found from the
present analysis($T_0=155$ MeV  and $c=1.88$), are much closer to
the values found from the analysis of pion production in $e^+e^-$
collisions\cite{Wilk7}($T_0=131 MeV$ and $c=1.83$); and $T_0 \sim
m_{\pi}$ can be recognized as the critical temperature of phase
transition of the hadronizing system in thermal
equilibrium\cite{Hagedorn2}. The reason of this discrepancy in the
outcome of the two studies could be attributed to the non-inclusion
of the constraints, given by eqn(22)-(23), in our previous analysis
involving charged pions and proton-antiprotons. Thus, it is
necessary to re-analyse the data on pion, proton and antiprotons
produced in RHIC energies in the light of the present model
including the constraints applied here, so that a conclusion could
be reached. Lastly, but more importantly, there is yet another very
interesting and important observation. The Fig.(2a) demonstrates
quite convincingly that the parameter $q$ does not seem to tend to
unity for even higher participant numbers, so non-extensivity may be
present even in the thermodynamical sense and not just as a
simulated microcanonical effect.
\par
We are quite aware that so many other topical issues of interest
around which we choose not to deal with in the present work as they
demand a separate treatment. The properties of energy dependence of
the multiplicity-density, inclusive cross-section and of the average
transverse momentum behaviour are just a few among the many. They
are both important and interesting as there are many reported
anomalies and puzzles\cite{Lykasov1,Lykasov2,Blok} which would be the subject of
matter of our future studies. In fact, we commented in just the
preceding paragraph that we would rework in the near future on the
pion, kaon and proton-antiproton spectra with the specific concerns
of constraining the parameters used by the method of calculation in
the non-extensive approach and try to throw light on the problems
and puzzles associated with the production characteristics of the
secondaries in the high energy collisions under studies.
\ \\
\ \\
{\bf{Acknowledgement:}} The authors are grateful to the learned
Referees for their valuable suggestions for betterment of an earlier
version of the present manuscript.\\
One of the authors, Bhaskar De(BD), acknowledges the financial
assistance received from Department of Science and Technology,
Government of India under the scheme SERC FAST TRACK PROPOSALS FOR
YOUNG SCIENTISTS 2006-2007(No. SR/FTP/PS-58/2005) .

\newpage
\begin{table}
\caption{Values of fitted parameters with respect to experimental
data on $\eta$-spectra produced in $P+P$ collision at $\sqrt{s_{NN}}=200$GeV}
\ \\
\centering
\begin{tabular}{|c|c|c|c|c|}
\hline  $N_{part}$ & $n_0$ & $q$ & $T_{eff}$(GeV) & $\chi^2/ndf$\\
\hline
2 & $0.075\pm0.003$  & $1.115\pm0.005$ & $0.097\pm0.006$ & $0.691/15$ \\
\hline
\end{tabular}
\end{table}

\begin{table}
\caption{Values of fitted parameters with respect to experimental
data on $\eta$-spectra at different centralities of $Au+Au$ and $D+Au$ collisions at RHIC}
\ \\
\centering
\begin{tabular}{|c|c|c|c|c|c|c|}
\hline  & Centrality & $N_{part}$ & $N$ & $c$ & $T_0$(GeV) & $\chi^2/ndf$\\
\hline
 & 0-20 & 279.9 & $24.8\pm0.1$  & $1.81\pm0.02$ & $0.165\pm0.003$ & $0.949/7$ \\
$Au+Au$  & 20-60 & 100.2 & $9.00\pm0.03$  & $1.90\pm0.03$ & $0.160\pm0.004$ & $0.268/7$ \\
 & 60-92 & 14.5 & $1.31\pm0.01$  & $1.82\pm0.04$ & $0.150\pm0.005$ & $0.313/3$ \\
 & Min. Bias & 109.1 & $9.70\pm0.03$  & $1.83\pm0.06$ & $0.162\pm0.004$ & $0.563/7$ \\
\hline
 & 0-20 & 15.6 & $1.43\pm0.01$  & $1.90\pm0.02$ & $0.160\pm0.004$ & $0.580/10$ \\
& 20-40 & 11.1 & $1.02\pm0.01$  & $1.92\pm0.03$ & $0.159\pm0.005$ & $0.553/10$ \\
$D+Au$ & 40-60 & 7.7 & $71\pm0.01$  & $1.88\pm0.03$ & $0.150\pm0.004$ & $0.237/10$ \\
 & 60-88 & 4.2 & $0.39\pm0.02$  & $1.91\pm0.02$ & $0.140\pm0.006$ & $1.657/9$ \\
 & Min. Bias & 9.1 & $0.84\pm0.04$  & $1.91\pm0.04$ & $0.151\pm0.001$ & $0.560/14$ \\
\hline
\end{tabular}
\end{table}

\begin{table}
\caption{Calculated values of $q$ and $T_{eff}$ for different participant nucleons($N_{part}$)}
\ \\
\centering
\begin{tabular}{|c|c|c|}
\hline  $N_{part}$ & $q$ & $T_{eff}$(GeV)\\
\hline
 2 & 1.115  & 0.097\\
\hline
 4.2 & 1.101  & 0.115\\
\hline
 7.7 & 1.096  & 0.123\\
\hline
 9.1 & 1.100  & 0.121\\
\hline
 11.1 & 1.097  & 0.130\\
\hline
 14.5 & 1.093  & 0.125\\
\hline
 15.6 & 1.096  & 0.131\\
\hline
 100.2 & 1.087  & 0.134\\
\hline
 109.1 & 1.086  & 0.137\\
\hline
 279.9 & 1.085  & 0.140\\
\hline
\end{tabular}
\end{table}

\newpage
\begin{figure}
\SetFigLayout{2}{2} \centering
\subfigure[]{\includegraphics[width=8cm]{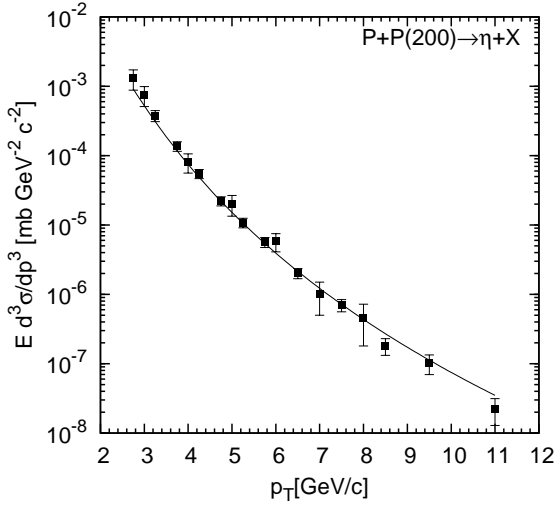}} \hfill
\subfigure[]{\includegraphics[width=8cm]{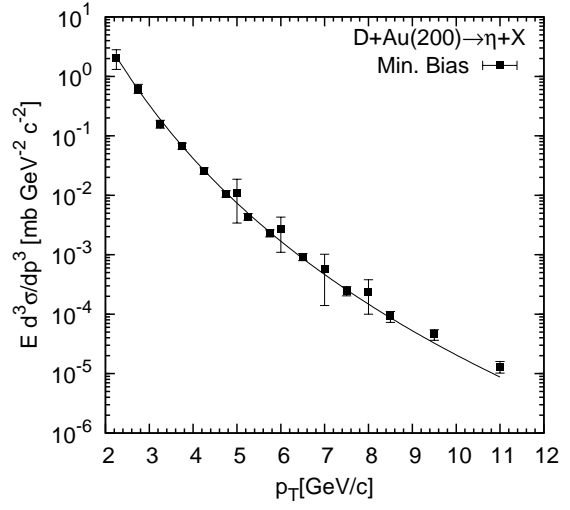}}\\
\subfigure[]{\includegraphics[width=8cm]{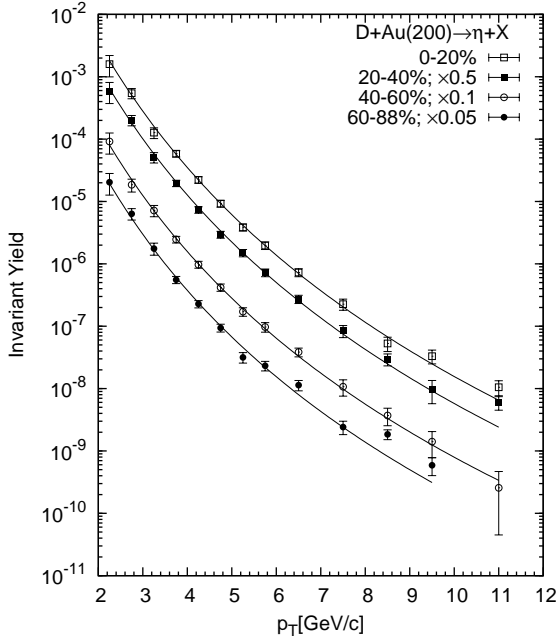}} \hfill
\subfigure[]{\includegraphics[width=8cm]{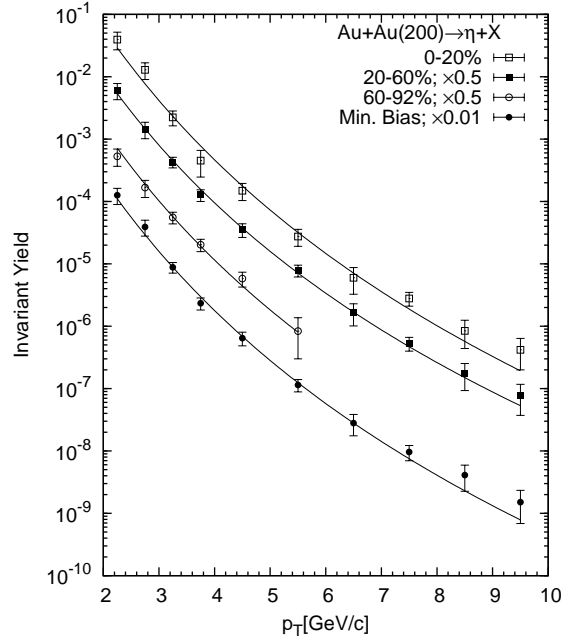}}
\caption{Plots of transverse momentum spectra of $\eta$-mesons
produced in $P+P$, $D+Au$ and $Au+Au$ collisions at $\sqrt{s_{NN}}=200$ GeV at different centralities. The filled symbols represent the experimental data
points\cite{Adler1}. The solid curves provide the fits
on the basis of nonextensive approach(eqn.(21)).}
\end{figure}

\begin{figure}
\SetFigLayout{1}{2} \centering
\subfigure[]{\includegraphics[width=8cm]{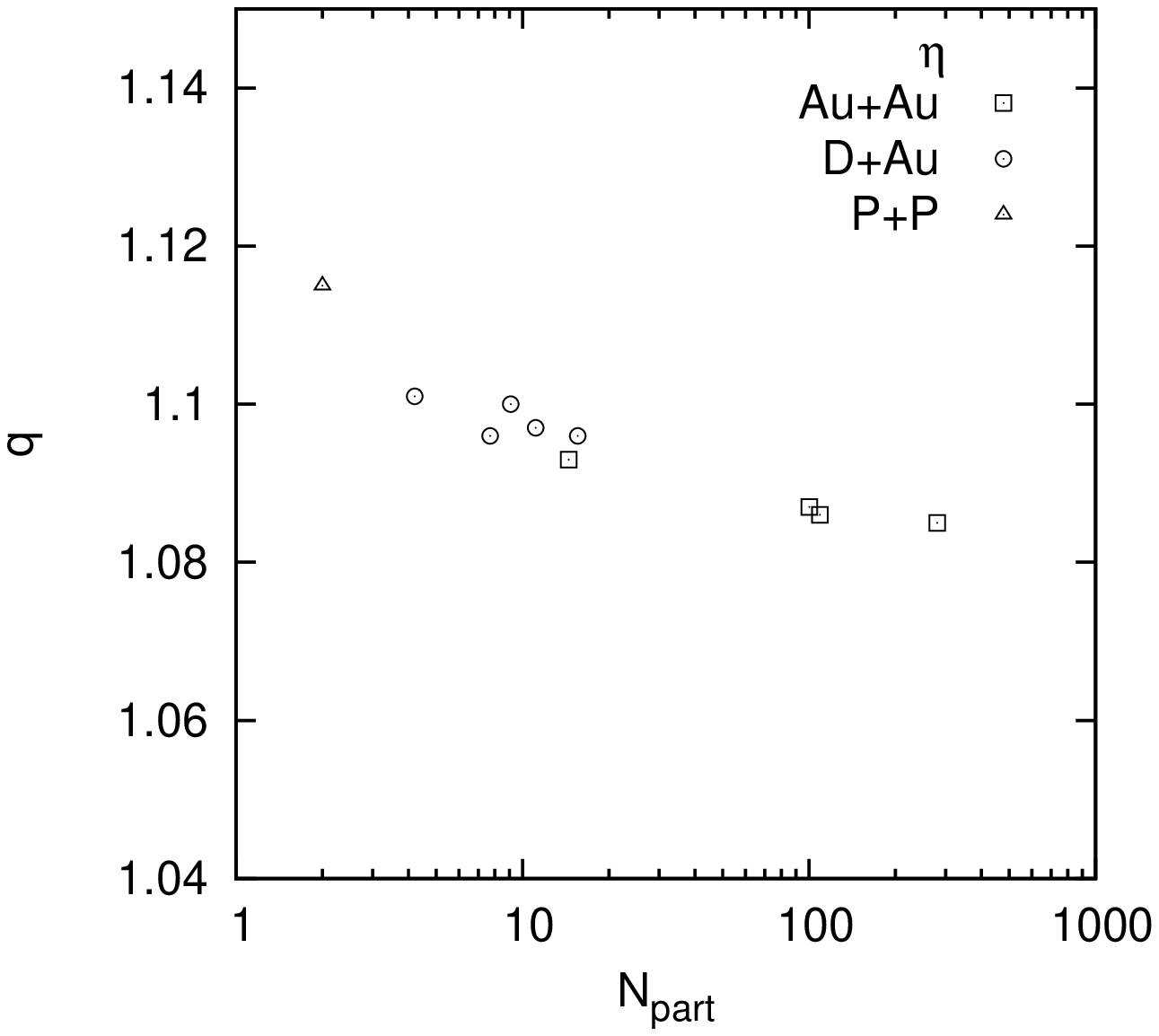}} \hfill
\subfigure[]{\includegraphics[width=8cm]{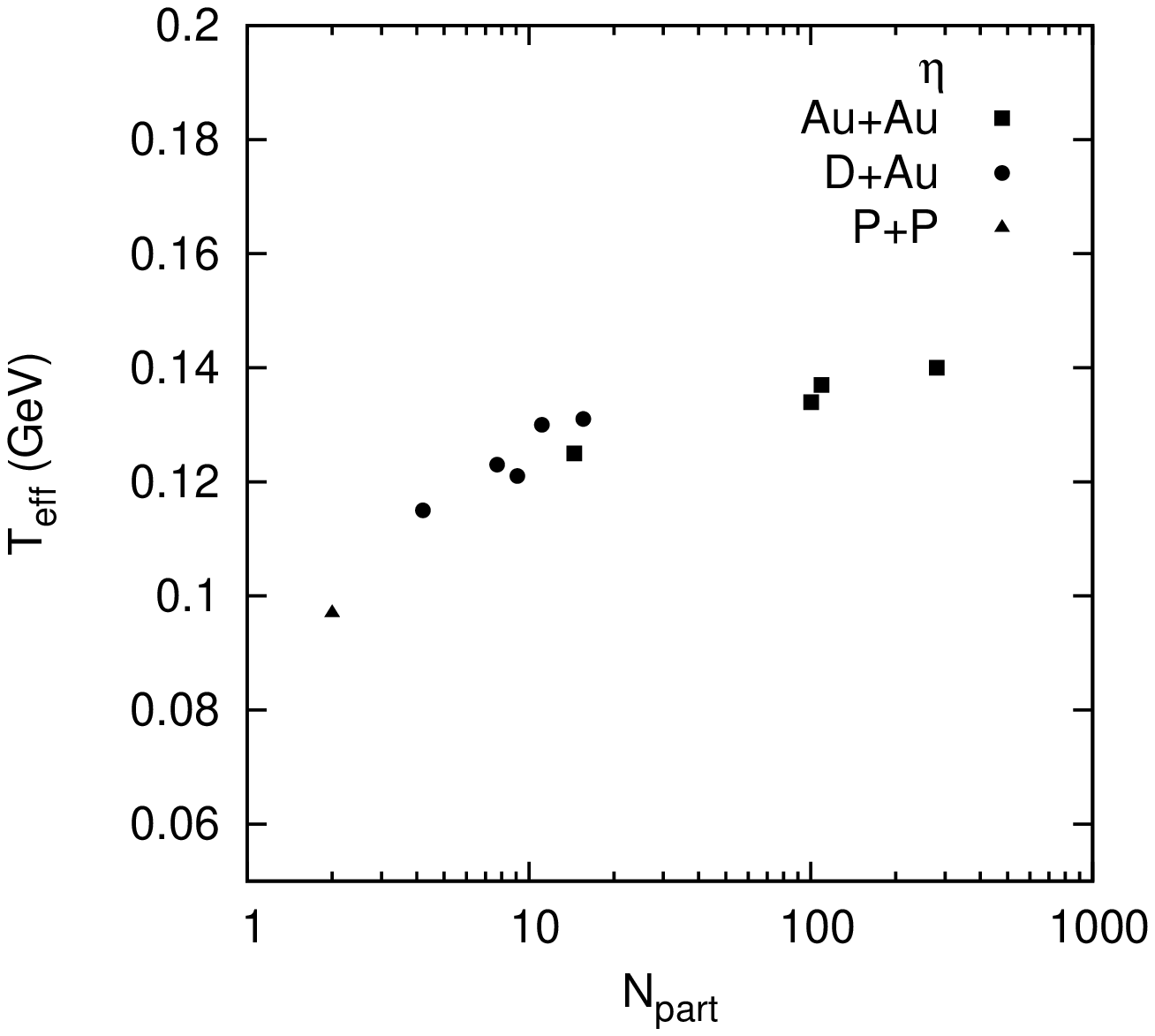}}
\caption{Plots of the nonextensive parameter $q$ and the effective temperature
$T_{eff}$ as a function of number of participant nucleons in $Au+Au$,
$D+Au$ and $P+P$ collisions at $\sqrt{s_{NN}}=200$ GeV for production of secondary $\eta$-mesons.}
\end{figure}

\begin{figure}
\SetFigLayout{1}{2} \centering
\includegraphics[width=8cm]{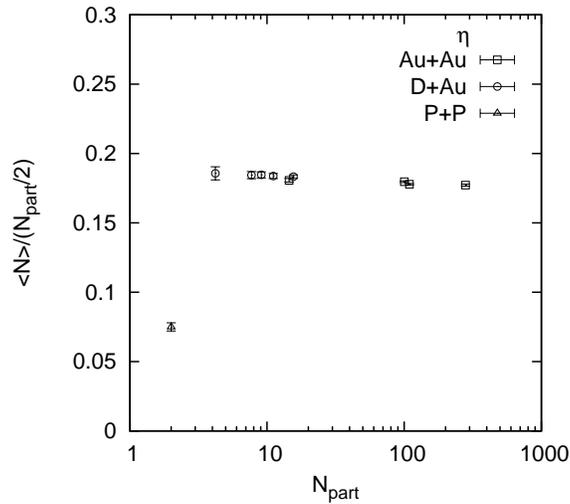}
\caption{Plots of the multiplicity of $\eta$-mesons produced per pair of participant nucleons
as a function of number of participant nucleons in $Au+Au$,
$D+Au$ and $P+P$ collisions at $\sqrt{s_{NN}}=200$ GeV.}
\end{figure}

\end{document}